\documentclass[a4paper, 12pt]{article}

\usepackage{fullpage}
\usepackage{authblk}
\usepackage{amsmath}
\usepackage{amssymb}
\usepackage{xfrac}
\usepackage{url}
\usepackage{csquotes}

\MakeOuterQuote{"}

\title{Solution for the Indefinite Integral of the \\ Standard Normal Probability Density Function}
\author{Joram Soch}
\affil{JoramSoch@web.de}
\date{December 15, 2015}

\begin{document}

\setcounter{page}{0}
\vspace*{1em}
\begin{center}
	\LARGE
	Solution for the Indefinite Integral of the \\ Standard Normal Probability Density Function \\ \vspace{1em}
	\large
	Joram Soch  \\ \vspace{1em}
	JoramSoch@web.de \\ \vspace{1em}
	December 15, 2015
\end{center}
\vspace*{1em}

\begin{abstract}
\noindent
Conventional wisdom assumes that the indefinite integral of the probability density function for the standard normal distribution cannot be expressed in finite elementary terms. While this is true, there is an expression for this anti-derivative in infinite elementary terms that, when being differentiated, directly yields the standard normal density function. We derive this function using infinite partial integration and review its relation to the cumulative distribution function for the standard normal distribution and the error function.
\end{abstract}

\vspace{1em}
\tableofcontents

\pagebreak
\section{Introduction}

Let $x \in \mathbb{R}$ be a random variable. $x$ is said to be \textit{normally distributed} with mean $\mu$ and variance $\sigma^2$, if its probability density function (PDF) is given by (Koch, 2007, eq.~2.166)

\begin{equation} \label{eq:N-pdf}
\varphi_{\mu,\sigma}(x) = \frac{1}{\sqrt{2 \pi} \sigma} \cdot e^{-\frac{1}{2} \left( \frac{x-\mu}{\sigma} \right)^2} \; ,
\end{equation}

such that its cumulative distribution function (CDF) is given by (Koch, 2007, eq.~2.168)

\begin{equation} \label{eq:N-cdf}
\Phi_{\mu,\sigma}(x) = \int_{-\infty}^{x} \varphi_{\mu,\sigma}(z) \, \mathrm{d}z \; .
\end{equation}

The \textit{standard normal distribution} is defined as the special case of the normal distribution with $\mu = 0$ and $\sigma^2 = 1$ and is characterized by the following PDF and CDF:

\begin{equation} \label{eq:SN-pdf-cdf}
\begin{split}
\varphi(x) &= \varphi_{0,1}(x) = \frac{1}{\sqrt{2 \pi}} \cdot e^{-\frac{1}{2} x^2} \; , \\
\Phi(x) &= \Phi_{0,1}(x) = \int_{-\infty}^{x} \varphi_{0,1}(z) \, \mathrm{d}z \; .
\end{split}
\end{equation}

Evaluating the definite integral $\Phi(x)$ requires knowing the indefinite integral of $\varphi(x)$. However, as can be proven by the Risch algorithm (Risch, 1969, 1970), there is no elementary function that solves \textit{Gaussian integrals} such as

\begin{equation} \label{eq:Gauss-Int}
\int e^{-x^2} \, \mathrm{d}x \; .
\end{equation}

In other words, this integral has no solution which is the composition of a finite number of arithmetic operations, exponentials, logarithms, constants and solutions of algebraic equations. Therefore, the \textit{error function} is defined as (Weisstein, 2015)

\begin{equation} \label{eq:erf}
\mathrm{erf}(x) = \frac{2}{\sqrt{\pi}} \int_{0}^{x} e^{-t^2} \, \mathrm{d}t
\end{equation}

and approximation methods such as numerical integration are used to evaluate $\mathrm{erf}(x)$ and $\Phi(x)$. In this work, we provide an analytical solution to the indefinite integral of $\varphi(x)$, i.e. a function $T(x)$ that satisfies the equations

\begin{equation} \label{eq:Tx}
\begin{split}
T(x) &= \int \varphi(x) \, \mathrm{d}x \\
\frac{\mathrm{d}}{\mathrm{d}x} T(x) &= \varphi(x)
\end{split}
\end{equation}

and use it to express $\Phi(x)$, $\Phi_{\mu,\sigma}(x)$ and $\mathrm{erf}(x)$.

\pagebreak
\section{Integration}

The indefinite integral of the standard normal PDF is given by

\begin{equation}
T(x) = \int \varphi(x) \, \mathrm{d}x = \int \frac{1}{\sqrt{2 \pi}} \cdot e^{-\frac{1}{2} x^2} \, \mathrm{d}x \; .
\end{equation}

The integrand can be rewritten as a trivial product, such that

\begin{equation}
T(x) = \frac{1}{\sqrt{2 \pi}} \int 1 \cdot e^{-\frac{1}{2} x^2} \, \mathrm{d}x \; .
\end{equation}

Applying the partial integration rule $\int u' v = u v - \int u v'$ yields

\begin{equation}
T(x) = \frac{1}{\sqrt{2 \pi}} \cdot \left[ x \cdot e^{-\frac{1}{2} x^2} + \int x^2 \cdot e^{-\frac{1}{2} x^2} \, \mathrm{d}x \right] \; .
\end{equation}

Applying partial integration to the remaining integral gives

\begin{equation}
T(x) = \frac{1}{\sqrt{2 \pi}} \cdot \left[ x \cdot e^{-\frac{1}{2} x^2} + \left[ \frac{1}{3} x^3 \cdot e^{-\frac{1}{2} x^2} + \int \frac{1}{3} x^4 \cdot e^{-\frac{1}{2} x^2} \, \mathrm{d}x \right] \right] \; .
\end{equation}

Another partial integration on the remaining integral gives

\begin{equation}
T(x) = \frac{1}{\sqrt{2 \pi}} \cdot \left[ x \cdot e^{-\frac{1}{2} x^2} + \left[ \frac{1}{3} x^3 \cdot e^{-\frac{1}{2} x^2} + \left[ \frac{1}{15} x^5 \cdot e^{-\frac{1}{2} x^2} + \int \frac{1}{15} x^6 \cdot e^{-\frac{1}{2} x^2} \, \mathrm{d}x \right] \right] \right] \; .
\end{equation}

We can generalize this result and conclude that

\begin{equation}
T(x) = \frac{1}{\sqrt{2 \pi}} \cdot \left[ \sum_{i=1}^{n} \left( \frac{x^{2i-1}}{(2i-1)!!} \cdot e^{-\frac{1}{2} x^2} \right) + \int \left( \frac{x^{2n}}{(2n-1)!!} \cdot e^{-\frac{1}{2} x^2} \right) \, \mathrm{d}x \right]
\end{equation}

where $(2n-1)!! = (2n-1) \cdot (2n-3) \cdot \ldots \cdot 3 \cdot 1$ is the double factorial (OEIS, A001147). This equation holds true for every $n \in \mathbb{N}^\ast$ and therefore also as $n$ tends to infinity:

\begin{equation}
T(x) = \frac{1}{\sqrt{2 \pi}} \cdot \left[ \sum_{i=1}^{\infty} \left( \frac{x^{2i-1}}{(2i-1)!!} \cdot e^{-\frac{1}{2} x^2} \right) + \lim_{n \to \infty} \int \left( \frac{x^{2n}}{(2n-1)!!} \cdot e^{-\frac{1}{2} x^2} \right) \, \mathrm{d}x \right] \; .
\end{equation}

Since $(2n-1)!!$ grows faster than $x^{2n}$, we will assume that

\begin{equation}
\lim_{n \to \infty} \int \left( \frac{x^{2n}}{(2n-1)!!} \cdot e^{-\frac{1}{2} x^2} \right) \, \mathrm{d}x = c \;
\end{equation}

for constant $c$ (proof not shown). This implies the indefinite integral

\begin{equation}
T(x) = \frac{1}{\sqrt{2 \pi}} \cdot e^{-\frac{1}{2} x^2} \cdot \sum_{i=1}^{\infty} \frac{x^{2i-1}}{(2i-1)!!} + c
\end{equation}

which means that $T(x)$ is just a product of $\varphi(x)$ and an infinite polynomial in $x$.

\pagebreak
\section{Differentiation}

The derivative $T'(x)$ of the function $T(x)$ is given by

\begin{equation}
T'(x) = \frac{\mathrm{d}}{\mathrm{d}x} \left( \frac{1}{\sqrt{2 \pi}} \cdot e^{-\frac{1}{2} x^2} \cdot \sum_{i=1}^{\infty} \frac{x^{2i-1}}{(2i-1)!!} + c \right) \; .
\end{equation}

Using the constant rule and sum rule, this can be rewritten as

\begin{equation}
T'(x) = \frac{1}{\sqrt{2 \pi}} \cdot \sum_{i=1}^{\infty} \frac{\mathrm{d}}{\mathrm{d}x} \left( \frac{x^{2i-1}}{(2i-1)!!} \cdot e^{-\frac{1}{2} x^2} \right) \; .
\end{equation}

Applying the product rule $(u v)' = u' v + u v'$ to the bracket yields

\begin{equation}
T'(x) = \frac{1}{\sqrt{2 \pi}} \cdot \sum_{i=1}^{\infty} \left[ \left( \frac{x^{2i-2}}{(2i-3)!!} \cdot e^{-\frac{1}{2} x^2} \right) - \left( \frac{x^{2i}}{(2i-1)!!} \cdot e^{-\frac{1}{2} x^2} \right) \right] \; .
\end{equation}

If we write out this sum for up to $i = 3$, we obtain

\begin{equation}
T'(x) = \frac{1}{\sqrt{2 \pi}} \cdot e^{-\frac{1}{2} x^2} \cdot \left[ \left( \frac{x^0}{(-1)!!} - \frac{x^2}{1!!} \right) + \left( \frac{x^2}{1!!} - \frac{x^4}{3!!} \right) + \left( \frac{x^4}{3!!} - \frac{x^6}{5!!} \right) + \ldots \right] \; .
\end{equation}

We can easily see that adjacent elements cancel out

\begin{equation}
- \frac{x^{2i}}{(2i-1)!!} + \frac{x^{2(i+1)-2}}{(2(i+1)-3)!!} = 0
\end{equation}

and therefore conclude that

\begin{equation}
T'(x) = \frac{1}{\sqrt{2 \pi}} \cdot e^{-\frac{1}{2} x^2} \cdot \left[ \frac{x^0}{(-1)!!} - \lim_{n \to \infty} \left( \frac{x^{2n}}{(2n-1)!!} \right) \right] \; .
\end{equation}

Again, since $(2n-1)!!$ grows faster than $x^{2n}$, we assume that

\begin{equation}
\lim_{n \to \infty} \left( \frac{x^{2n}}{(2n-1)!!} \right) = 0
\end{equation}

(proof not shown). By definition, it also holds that

\begin{equation}
\frac{x^0}{(-1)!!} = 1 \; .
\end{equation}

Taken together, this implies the derivative

\begin{equation}
T'(x) = \frac{1}{\sqrt{2 \pi}} \cdot e^{-\frac{1}{2} x^2} = \varphi(x)
\end{equation}

which proves that $T(x)$ is the anti-derivative of $\varphi(x)$.

\pagebreak
\section{Application}

\subsection{Normal CDFs}

We can easily relate the anti-derivative $T(x)$ to the CDF $\Phi(x)$ by adjusting the constant $c$ in $T(x)$. Since $\Phi(0) = 0.5$ and $T(0) = c$, it follows that $c_0 = 0.5$ and $\Phi(x)$ is given by

\begin{equation} \label{eq:SN-cdf-T}
\Phi(x) = \varphi(x) \cdot \sum_{i=1}^{\infty} \frac{x^{2i-1}}{(2i-1)!!} + 0.5 \quad \text{or} \quad \Phi(x) = T(x) - c + 0.5 \; .
\end{equation}

This is equivalent to an equation given by George Marsaglia (Marsaglia, 2004, p.~5) which however was derived using Taylor series expansion at $x = 0$.

\vspace{1em}

The CDF of the standard normal distribution $\mathcal{N}(0,1)$ can be related to $\Phi_{\mu,\sigma}(x)$ by

\begin{equation} \label{eq:N-SN-cdf}
\Phi_{\mu,\sigma}(x) = \Phi \left( \frac{x-\mu}{\sigma} \right) \; .
\end{equation}

This entails that the CDF for the general normal distribution $\mathcal{N}(\mu,\sigma)$ is given by

\begin{equation} \label{eq:N-cdf-T}
\Phi_{\mu,\sigma}(x) = \varphi(z) \cdot \sum_{i=1}^{\infty} \frac{z^{2i-1}}{(2i-1)!!} + 0.5 \quad \text{where} \quad z = \frac{x-\mu}{\sigma} \; .
\end{equation}

\subsection{Error Function}

As stated in equation (\ref{eq:erf}), the error function is defined as

\begin{equation} \label{eq:erf-def}
\mathrm{erf}(x) = \frac{2}{\sqrt{\pi}} \int_{0}^{x} e^{-t^2} \, \mathrm{d}t
\end{equation}

which means that the error function is twice the definite integral of the normal density with mean 0 and variance $\sfrac{1}{2}$, minus 1:

\begin{equation} \label{eq:erf-Phi}
\mathrm{erf}(x) = 2 \cdot \Phi_{0,\sqrt{\sfrac{1}{2}}}(x) - 1 = 2 \cdot \Phi \left( \sqrt{2} x \right) - 1 \; .
\end{equation}

If we combine this with equation (\ref{eq:SN-cdf-T}), we obtain:

\begin{equation} \label{eq:erf-T}
\mathrm{erf}(x) = \frac{2}{\sqrt{\pi}} \cdot e^{-x^2} \cdot \sum_{i=1}^{\infty} \left( 2^{i-1} \cdot \frac{x^{2i-1}}{(2i-1)!!} \right) \; .
\end{equation}

This is equivalent to an equation given by Forman S. Acton (Acton, 1990, eq.~1.12) which however was derived by solving a differential equation system. The formula is also listed by Eric W. Weisstein (Weisstein, 2015, eq.~10) who suggests to employ it for computation of $\mathrm{erf}(x)$ when $x \ll 1$.

\pagebreak
We conclude this section by comparing formula (\ref{eq:erf-T}) with MATLAB's \verb|erf| and \verb|normcdf| for computation of $\mathrm{erf}(x)$. As we cannot compute an infinite number of terms, we replace $\infty$ by $n$ in equation (\ref{eq:erf-T}) and vary $n$ from 2 to 20 in steps of 2. All functions are calculated in the interval $X = [-2.5,+2.5]$. Results are given in Figure~1.

In summary, \verb|erf| and \verb|normcdf| deliver identical results and the $T(x)$ approach gives acceptable results on $X$ for $n \geq 14$. With $n = 14$, the relative error compared to $\mathrm{erf}(x)$ is at most 0.33\%. For higher $|x|$, higher $n$ is needed for exact results. However, $|\mathrm{erf}(x)| \approx 1$ for $|x| > 2.5$, so that $n > 20$ will not be required in practice.

\vspace{1em}
\begin{center}
\includegraphics[width=1.0\linewidth, clip=true, trim=100 40 100 20]{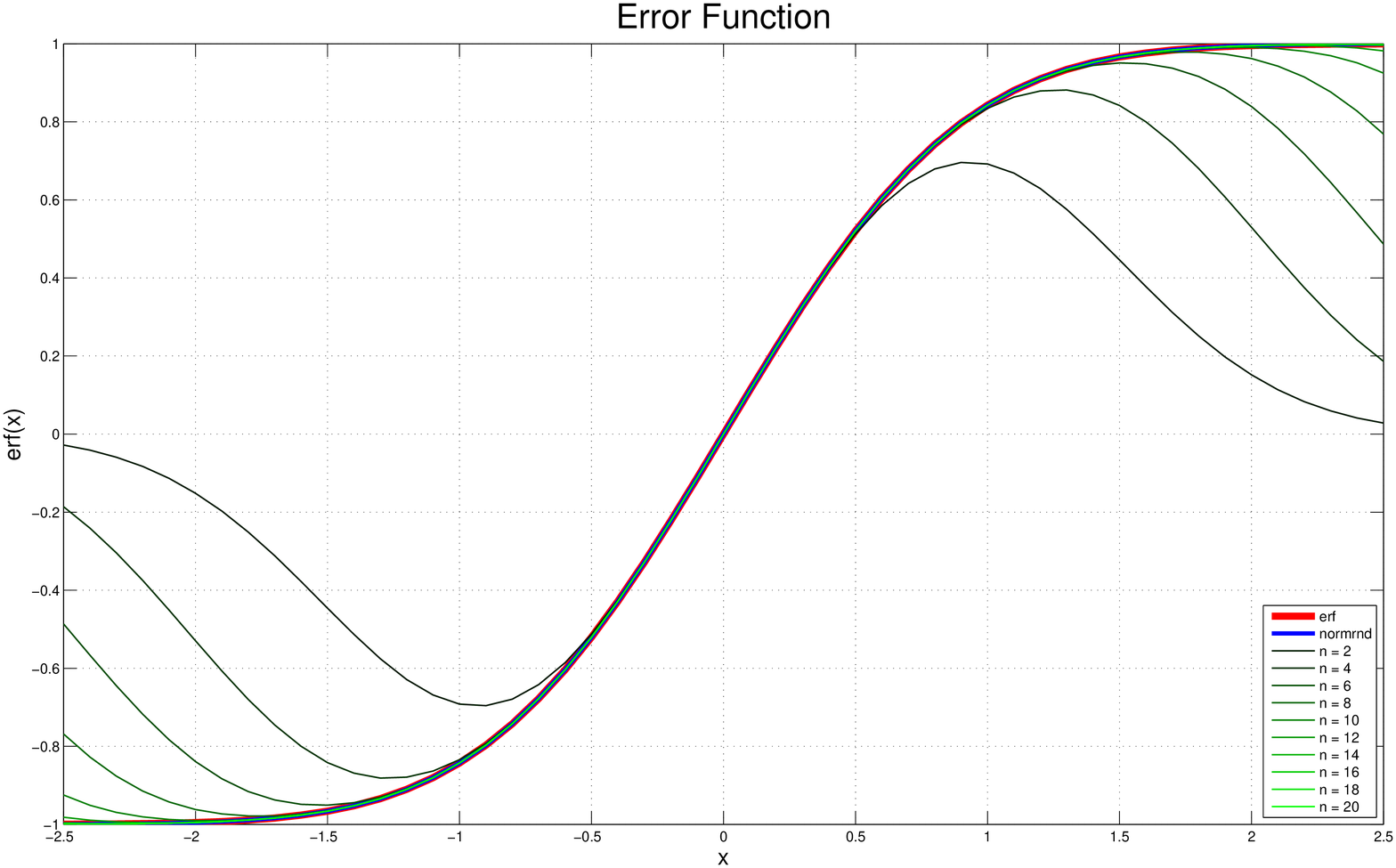}
\end{center}

\textbf{Figure 1.} Comparison of error function implementations. The error function $\mathrm{erf}(x)$ was calculated using MATLAB's \verb|erf| (red), using MATLAB's \verb|normcdf| and equation (\ref{eq:erf-Phi}) (blue) and using equation (\ref{eq:erf-T}) (green). The first two approaches produce the same results. The more terms are calculated for $T(x)$, the better it approximates $\mathrm{erf}(x)$. On the interval $X = [-2.5,+2.5]$, $n \geq 14$ is enough for practical purposes. All computations were performed using MATLAB R2013b.

\pagebreak
\section{Conclusion}

Using infinite partial integration, we have derived an analytical expression for the indefinite integral of the standard normal PDF

\begin{equation}
T(x) = \int \frac{1}{\sqrt{2 \pi}} \cdot e^{-\frac{1}{2} x^2} \, \mathrm{d}x
\end{equation}

and linked it to previous results for the error function (Acton, 1990) as well as the general and standard normal CDF (Marsaglia, 2004)

\begin{equation}
\Phi(x) = \int_{-\infty}^{x} \frac{1}{\sqrt{2 \pi}} \cdot e^{-\frac{1}{2} z^2} \, \mathrm{d}z \; .
\end{equation}

Cursorily calculations demonstrate that relatively few terms of the infinite sum have to be computed for satisfying accuracy.

\section{References}

\renewcommand{\section}[2]{}


\begin{thebibliography}{9}

\bibitem{Acton_1990}
Acton FS (1990): \textit{Numerical Methods that (usually) work}. Mathematical Association of America, Washington, D.C.

\bibitem{Koch_2007}
Koch KR (2007): \textit{Introduction to Bayesian Statistics}. 2\textsuperscript{nd} Edition, Springer, Berlin/\linebreak[2]Heidelberg.

\bibitem{Marsaglia_2004}
Marsaglia G (2004): "Evaluating the Normal Distribution". \textit{Journal of Statistical Software}, vol.~11, iss.~4, pp.~1-11.

\bibitem{Risch_1969}
Risch RH (1969): "The Problem of Integration in Finite Terms". \textit{Transactions of the American Mathematical Society}, vol.~139, pp.~167-189.

\bibitem{Risch_1970}
Risch RH (1970): "The Solution of the Problem of Integration in Finite Terms". \textit{Bulletin of the American Mathematical Society}, vol.~76, no.~3, pp.~605-608.

\bibitem{Weisstein_2015}
Weisstein EW (2015): "Erf". \textit{MathWorld, a Wolfram Web Resource}, last retrieved on Dec 15, 2015; URL: \url{http://mathworld.wolfram.com/Erf.html}.

\bibitem{OEIS}
\textit{The On-Line Encyclopedia of Integer Sequences} (OEIS); URL: \url{http://oeis.org/}.

\end{thebibliography}
\end{document}